\documentclass[aps,prb,amsmath,amssymb,twocolumn,letterpaper,%
         footinbib,bibnotes,showpacs,reprint,balancelastpage,raggedbottom,%longbibliography,%
         citeautoscript,floatfix]{revtex4-1}
\usepackage[usenames,dvipsnames,table]{xcolor}
\usepackage[pdftex,
  breaklinks,             %
  bookmarks = false, %
  pdfpagemode   = UseNone,%
  pdfstartview  = FitH,     %
  pdfstartpage  = 1,      %
  colorlinks    = true,   %
]{hyperref}
\hypersetup{
    linkcolor=RubineRed,          % color of internal links
    citecolor=ForestGreen,        % color of links to bibliography
    filecolor=Mulberry,      % color of file links
    urlcolor=RoyalBlue           % color of external links
}
\usepackage{amssymb}
\usepackage{graphicx}
\usepackage{float}
\usepackage{enumerate}
\usepackage{microtype}
\usepackage{xspace}



\begin{document}

\author{Liang-Feng Huang}
  \affiliation{Department of  Materials  Science  and  Engineering,  Northwestern  University,  Evanston,  IL  60208,  USA}
\author{James M.\ Rondinelli}
\email{jrondinelli@northwestern.edu}
  \affiliation{Department of  Materials  Science  and  Engineering,  Northwestern  University,  Evanston,  IL  60208,  USA}

\title{Accurate First-Principles Electrochemical Phase Diagrams for Ti Oxides from Density Functional Calculations}

\begin{abstract}
Developing an accurate simulation method for the electrochemical stability of solids, as well
as understanding the physics related with its accuracy, is critically important for improving 
the performance of compounds and predicting the stability of new materials in aqueous environments. 
Herein we propose a workflow for the accurate 
calculation of first-principles electrochemical phase (Pourbaix) diagrams. 
With this scheme, we study the electrochemical stabilities of Ti and Ti oxides using density-functional theory. 
First, we find the accuracy of an exchange-correlation functional in predicting 
formation energies and electrochemical stabilities is closely related
with the electronic exchange interaction therein. 
Second, the 
\textcolor{black}{metaGGA} 
and hybrid functionals with a more precise description 
of the electronic exchange interaction lead to a systematic improvement in 
the accuracy of the Pourbaix diagrams. 
Furthermore, we show that accurate Ti Pourbaix diagrams also require that thermal effects are 
included through vibrational contributions to the free energy.
We then use these diagrams to explain various experimental electrochemical phenomena for 
the Ti--O system, and show that if  experimental formation energies for Ti oxides, 
which contain contributions from defects  owing to their generation at high (combustion) temperatures,
are directly used to predict room temperature Pourbaix diagrams then significant
\textcolor{black}{inaccuracies result.}
In contrast, the formation energies from accurate first-principles calculations, \emph{e.g.}, using 
\textcolor{black}{metaGGA} 
and hybrid functionals, are found to be more reliable. Finally to facilitate the future application of our accurate electrochemical phase equilibria diagrams, 
the variation of the Ti Pourbaix diagrams with aqueous ion concentration is also provided. 
\end{abstract}


\maketitle

\section{Introduction}
The phase stability of transition metals and their native complex
oxides under variable pH and potential bias is a foundational
concept utilized for the design of functional materials in a variety
of industries ranging from high-performing corrosion resistant
alloys,\cite{Davis2000} superior electrode materials for energy
storage systems,\cite{Saha2014,Kanevskii2005} and biocompatibility
of implants.\cite{Davis2000,Geetha54}
This fact is clear when considering  Ti, which beyond being
an important structural material owing to its excellent mechanical properties, has high
corrosion resistance and high biocompatiblity, is light weight, and
readily alloys with others elements.
\cite{Leyens2003,Lutjering2007,Banerjee61,Geetha54,Niinomi33A,Yuan2014}
These features make it well suited for applications in
aqueous and \emph{in vivo} environments.
Moreover, passivating Ti oxides enhance the corrosion
resistance of metallic Ti \cite{Kelly1982}, whereas titanate oxides such as TiO$_2$
have applications in pigmentation, sensors, solar cells, and photocatalysis
(\emph{e.g.}, water splitting and pollutant treatment),
derived from the oxides superior electronic and optical properties
\cite{Grant1959,Mo1995,Lazzeri2001,Muscat2002,Janotti2010,Hanaor2011}.
Nonetheless, the operational materials performance of Ti and its native
oxides are in part limited by their electrochemical properties, specifically corrosion
susceptibility, which can be understood from solid-aqueous phase equilibria
diagrams.

The main materials chemistry tool to understand such thermodynamics includes the
electrochemical Pourbaix diagram,\cite{Pourbaix1966} which maps the response
of pH and potential ($V$) on the phase stability of the metal, oxides,
and/or hydroxides and ions in solution.
Pourbaix diagrams may be directly generated by detailed and challenging
electrochemical measurements.\cite{Yang1997,Hinai2014} 
Although it is often difficult 
to achieve a reliable Pourbaix diagram in one set (or several sets) of experimental measurements, 
because of challenges in controlled sample preparation and accurate characterization.\cite{kelly2002electrochemical} 
Alternatively, as is more common, Pourbaix diagrams are computed by using 
(a) the collected experimental formation energies\cite{Beverskog1996,Beverskog1997,Beverskog1997_2,Portero2011} and/or 
(b) the calculated formation energies,\cite{Persson2012,Zeng2015} which are described in more detail below. 
However, both experimental and theoretical formation 
energies are sometimes prone to inaccuracy owing to various causes, and the computed Pourbaix diagram may not be consistent with direct electrochemical measurements. 
Computational methods are also particularly sought
for cases where new alloys and oxides
are being developed, \emph{i.e.}, to enable predictive materials design, or for rapid phase space exploration as in cases
 where the potential phase space of candidate oxides and hydroxides is sufficiently large to make
discerning various phases from experiment impractical.
The first approach relies on experimental chemical potentials of
aqueous ions at \emph{room temperature} and the formation energies of
oxides estimated from \emph{high-temperature} combustion reactions.
This information is then used as input into the relevant
thermo-electrochemical equations to generate the
solid-aqueous phase equilibria diagram.
This approach, for example, was previously used to predict the Ti
Pourbaix diagram, however, the resulting stability of various phases
were found to be inconsistent with direct electrochemical
measurements: TiO$_2$ was predicted to corrode under acid
solutions with pH values of $\lesssim$-1.0
\cite{Pourbaix1966,Portero2011,Devic2014}, while various
electrochemical measurements show that the corrosion boundary
of TiO$_2$ is at a pH value of $0.8\sim2.0$
\cite{Kelly1982,Yin2001,*Kroger2006,*Hu2009,*Yigit2009,*Lee2014}.
Furthermore, TiO appears at lower potentials than Ti$_2$O$_3$ in many Pourbaix
diagrams simulated using this approach,\cite{Pourbaix1966,Portero2011,Persson2012} whereas TiO is not observed in electrochemical measurements \cite{Kelly1982}.
Such inconsistencies may be reconciled by understanding that the
experimental formation energies for oxides are estimated from heat
of combustion reactions at high temperatures \cite{Chase1998}, where
a high concentration of stoichiometric/nonstoichiometric defects are
readily formed \cite{Grant1959,Murray1987,Bartkowski1997}, and are
known to significantly affect the oxide stability
\cite{Murray1987,*Leung1996,*Bartkowski1997}.
Accurately predicting the electrochemical stability of transition
metals and oxides, such as Ti and Ti oxides at low (room)
temperature, therefore, indicates that available experimental oxide
formation energies obtained from high temperature measurements
should probably be avoided. %%

The second approach averts this complication by utilizing
first-principles calculations based on density functional theory (DFT)
to compute the formation energies of the
pristine oxides and combining the ab-initio energies with
experimental chemical potentials for ions in solution.\cite{Persson2012,Zeng2015}
However, this introduces a different challenge: How accurate are the
calculated formation energies of transition metals and their oxides?
This accuracy depends on the the ability of the exchange-correlation potential ($V_{xc}$)
within the DFT formalism to produce the detailed balance of an
ionocovalent metal--oxygen bond and to gives energies for reference state species
that are close to experimental values.
Within this context, the
local density approximation (LDA) \cite{Ceperley1980,Perdew1981} and
generalized gradient approximation (GGA)
\cite{Perdew1996,Perdew1997} are the two most frequently used semilocal
density functionals, and their inaccuracies are well documented regarding the
famous over-binding problem in the O$_2$ molecule
\cite{Jones1989,Adllan2011} and the self-interaction error
in transition-metal oxides \cite{Jones1989,Wang2006}.

To remove the inaccuracy of these functionals, various `fitting' corrections
\cite{Wang2006,Lany2008,Jain2011,Lutfalla2011,Stevanovic2012,Aykol2014}
to the energies of the reference species (O$_2$ and metal) or to the
electronic potential for the oxides (Hubbard plus $U$ correction
\cite{Anisimov1991,*Liechtenstein1995,*Dudarev1998,*Anisimov1997,*Bengone2000,*Rohrbach2003})
are used to improve the agreement with experiment.
%
%\jmr{what kind of experiments are being compared to here, not the ones mentioned in the first approach, is that correct?}
%
Although these numerical corrections may be useful
when fast calculations are required, \emph{e.g.}, in large-scale
structural searches and high-throughput simulations, where relatively modest
accuracies in total energies are acceptable for a meaningful result \cite{Hautier2010,Jain2011_2}, they may result in
an unphysically large $U$ value.
This is the case in titanate oxides \cite{Aykol2014,Morgan2007,*Morgan2010,*Hu2011},
where values from $4\le U \le 9$\,eV are used for the $3d$ manifold, despite correlations
typically being weak owing to the small valence orbital occupancy 
\cite{Mo1995,Morgan2007,Dompablo2011,Jiang2013_2}.
\textcolor{black}{Furthermore, the experimental formation energies for which the
DFT energies are corrected to achieve agreement may
also introduce some additional inaccuracy owing to the contribution
of the defect stabilizing effect in the experiments.}
%Furthermore, the experimental formation energies for which the
%DFT energies of stoichiometric phases are corrected to achieve agreement may
%also introduce some additional inaccuracy owing to the contribution
%of non-stoichiometric phases present in the experiments.
%%

%%
On the other hand, exchange-correlation functionals
with a more sophisticated treatment of the \emph{electronic exchange interaction},
through for example adding nonlocal electronic exchange (in metaGGA and hybrid functionals) 
to that of the homogeneous electron gas, can improve the thermodynamic stability, lattice constants,
electronic properties, and phonon frequencies of transition metals and
related compounds.
\cite{Muscat2002,Paier2008,*Chevrier2010,*Janotti2010,*Dompablo2011,*Hu2011,*Yan2013,*Corso2013,*Aras2014,*Medasani2015}
In transition metal oxides, the localized $d$ orbital renders the electron density highly inhomogeneous and environment dependent, which results in an obvious nonlocal character of the electronic Hamiltonian, especially for the electronic exchange. 
This electronic exchange nonlocality is uniquely treated in both 
metaGGA and hybrid functionals, \emph{i.e.}, by using an additional electronic kinetic energy term and nonlocal Fock exchange, respectively.  
Such functionals could provide a more systematic path towards
realizing meaningful predictions of electrochemical behavior.
Nonetheless, the extent to which these functionals are better for
simulating Pourbaix diagrams and the underlying mechanism
governing their accuracy has yet to be reported.

Here we describe a streamlined first-principles workflow that combines DFT calculations without \emph{ad hoc} corrections for the formation energies of solids with experimental chemical potentials of aqueous ions at standard state.
We apply this scheme to study Ti and Ti oxides, focusing first on the
functional dependence of these compounds to explain the
connection between the electronic exchange interaction
and the electronic structure.
With this understanding, we justify the
exchange-correlation functional dependence of the DFT calculated Ti Pourbaix diagrams
and explain various experimental electrochemical
phenomena with our accurate diagrams that include portions of exact-Fock exchange.
Lastly, we examine the variation of the diagrams with aqueous ion concentration
and inclusion of lattice vibrational factors in our simulations.
Our work provides a framework for accurate ab-initio simulations
of transition-metal Pourbaix diagrams under different environmental conditions and
provides updated and reliable Ti Pourbaix diagrams that are likely to be useful for both
scientific research and industrial technologies.
\section{Materials and Theoretical Methods}\label{methods}

\subsection{Titanium Materials System}
The considered Ti oxides include: Ti$_6$O, Ti$_3$O, Ti$_2$O,
Ti$_3$O$_2$, TiO, Ti$_2$O$_3$, Ti$_3$O$_5$, Ti$_4$O$_7$, TiO$_2$
(rutile), and Ti$_2$O$_7$; the former four of which are oxides with
interstitial O atoms whereas some of the others with Ti$_n$O$_{2n-1}$
stoichiometries correspond to the Magn\'eli phases.
Except for Ti$_3$O$_2$ and Ti$_2$O$_7$, all
structures are obtained from the {\it{Inorganic Crystal
Structure Database}} \cite{ICSD}.
Ti$_3$O$_2$ is obtained by placing one more O atom into Ti$_3$O from which
we find from our DFT calculations (detailed next) that an
$AB$-type stacking arrangement of interstitial O atoms is the most stable configuration.
Ti$_2$O$_7$ is obtained from the {\it{Materials
Project}} \cite{Jain2013}.
The structures obtained from these databases are used as the initial
input for our simulations; the structural details of each 
are given in Ref.~\onlinecite{Supplmental_Note:PRB}.

Note that more than 100 titanate oxide (Ti$_m$O$_n$) variants
are reported experimentally \cite{Murray1987} with variable oxygen content
$x=n/(m+n)$ ranging from 0 to $\frac{2}{3}$.
Here, we have selected the aforementioned compounds as representative of such
phases, and  those examined have oxygen concentrations spanning the same oxygen
content range.
The additional phases may be found in the
{\it{Inorganic Crystal Structure Database}}\cite{ICSD},
{\it{Materials Project}} \cite{Jain2013}, and
{\it{Open Quantum Materials Database}} \cite{Saal2013}.

\subsection{Computational Methods}

\par
\begin{figure}
\centering
\scalebox{1}[1]{\includegraphics{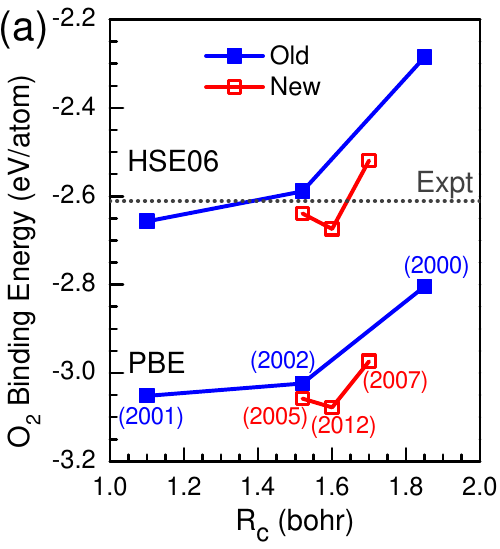}}
\scalebox{1}[1]{\includegraphics{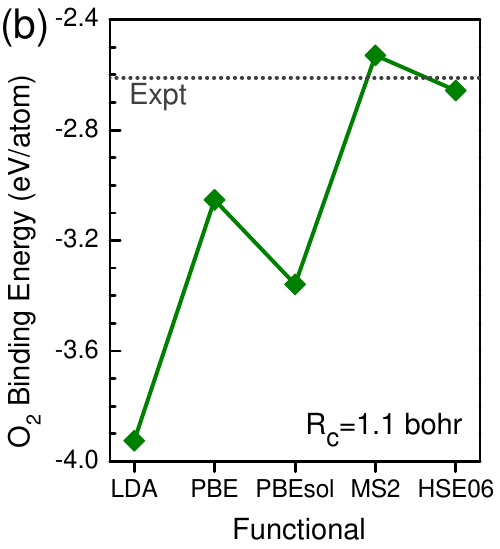}}\vspace{-0.75\baselineskip}
\caption{\label{Fig_O2_Binding_Energy} The dependence of the O$_2$
binding energy on the type of (a) pseudopotential and (b) density
functional at $R_c=1.1\,\mathrm{bohr}$, where $R_c$ is the ion radius in the PAW pseudopotential. The
zero-point energy correction (-0.049 eV) has been used in the
experimental value (-2.611 eV). In panel (a), the generation years
of the pseudopotentials provided by VASP
are indicated in parentheses.
}
\end{figure}

 The structures and energies are calculated using DFT \cite{Martin2004},
which is implemented in the Vienna \emph{Ab Initio} Package (VASP).\cite{Kresse1996,Kresse1996_2,Hafner2008}
The interaction among the core and valence electrons is treated using the projector augmented-wave (PAW) method \cite{Blochl1994,Kresse1999}.
The details of this separation of electrons may influence the formation energies
involved in calculating the electrochemical phase diagrams.
Indeed, \autoref{Fig_O2_Binding_Energy}(a) shows
how the O$_2$ binding energy [$E_b$(O$_2$)] depends
on the PAW ion-core radius ($R_c$).
By comparing our results with the hardest PAW pseudopotential ($R_c=1.1$ bohr) used
for different functionals, we find that PAWs with $R_c>1.6$ bohr have
a lower accuracy.
For this reason, the hardest PAW pseudopotentials for the O
(atomic configuration $2s^22p^4$; $R_c=1.1$ bohr) and
Ti ($3s^23p^63d^24s^2$; 2.3 bohr) atoms are used
here, and a cutoff energy of 900 eV is used in the planewave expansion to
achieve an energetic convergence of $<3$ meV per atom.
The reciprocal-space $k$-grid density is set to be
$\gtrsim\frac{26}{a_0}\times\frac{26}{b_0}\times\frac{26}{c_0}$ for
the calculation of Ti oxides, where $a_0$, $b_0$, and $c_0$ are
lattice constants of the periodic cell 
\textcolor{black}{scaled by the unit of angstrom}.
For pure metallic Ti, the grid is 1.5 times denser along each direction.
The convergence threshold for the self-consistency electronic iteration
is set to be $10^{-6}$\,eV, which is required for accurate forces, stresses, and
vibrational frequencies. The atomic positions, cell volume, and cell shape are
globally optimized, where the convergence thresholds for energy,
force, and stress are set to be $10^{-5}$ eV, $10^{-3}$ eV/\AA{},
and $0.2$ GPa, respectively.

To study the effect of electronic
exchange, exchange-correlation functionals with different approximations along
Jacob's ladder\cite{Perdew_Jacob} are used and compared: (i) local density
approximation (LDA), (ii) generalized gradient approximation (GGA),
(iii) 
\textcolor{black}{metaGGA}, 
and (iv) hybrid functional. The LDA functional used here is parametrized by Perdew and Zunger,
\cite{Ceperley1980,*Perdew1981} and the exchange-correlation interaction at a local site equals 
that of a homogeneous electron gas with the same electron density.
Two functionals at the GGA level are used,
\emph{i.e.}, PBE \cite{Perdew1996,*Perdew1997} and PBEsol
\cite{Perdew2008,*Perdew2009}. PBE considers the electron-density inhomogeneity by 
including the density gradient, and PBEsol is derived by making the PBE exchange enhancement 
factor have the same asymptotic trend as the fully homogeneous gas given by the LDA.
\textcolor{black}{
Note that the PBE functional is used to construct the PAWs for all DFT calculations that use 
the PBEsol, MS2, and HSE06 functionals, whereas the PAWs used in the LDA calculations are generated using the LDA functional.
}

Various functionals at the 
\textcolor{black}{metaGGA} 
level include TPSS \cite{Tao2003}, revTPSS \cite{Perdew2009_2}, and MS2 \cite{Sun2013,*Sun2013_2}, which mainly 
capture the nonlocal electronic exchange by additionally including electronic kinetic energy.
Here we employ the MS2 potential owing to its higher numerical stability
found in our benchmark calculations.
Finally, the HSE06 hybrid functional \cite{Heyd2003,*Heyd2004,*Heyd2004_2,*Heyd2006,*Vydrov2006} is used to
study the effect of nonlocal electronic exchange, where 25\,\% semilocal PBE electronic exchange is replaced 
by the exact Fock exchange, and a screening length of 10\,\AA\ is used for numerical convergence.

Note that no corrections to the formation energies of the oxides are
made in this work, because such values can be obtained by using
more accurate functionals which include sophisticated treatments of
exact electronic exchange (described in detail in the Results section).
For example, the O$_2$ binding is largely overestimated by the LDA,
PBE, and PBEsol functionals [\autoref{Fig_O2_Binding_Energy}(b)],
with the PBEsol $E_b$(O$_2$) intermediate between the others,
because the electronic exchange in PBEsol is
intermediate between LDA and PBE.\cite{Perdew2008,Perdew2009}
However, the MS2, and HSE06 functionals with nonlocal
electronic exchange interaction have an accuracy of
$<\!0.07$ eV for $E_b$(O$_2$) with respect to the
experimental result. For this reason, no artificial correction is
necessary for oxide $E_f$ when functionals with more precise electronic exchange
are included.

The phonon spectra are calculated using the
small-displacement \textcolor{black}{(or frozen phonon)} method
\cite{Kresse1995,*Parlinski1997}, which is implemented in the PHONOPY code
\cite{Togo2008}. The Ti and titanate oxide supercells used for the
phonon calculations \textcolor{black}{and a description of the relative performance of 
each functional in reproducing the experimental crystal structure} are provided in Ref.~\onlinecite{Supplmental_Note:PRB}.
To obtain accurate phonon band dispersions, there are two criteria that
we have tried to fulfill: (1) the supercell lattice constants should be
$>9$ \AA{} and (2) both the Brillouin-zone center and boundary of the
unit cell should be sampled by the supercell, \emph{i.e.}, the supercell
lattice constants should be an even multiple of those of the
unit cell.
When these two criteria are not fulfilled,
artificial imaginary vibrational modes sometimes (not always) appear.

\textcolor{black}{
Although the overbinding character of the  LDA functional leads to some minor 
discrepancies in the lattice constant compared to experiment, \textcolor{black}{ the free energy always has quite small functional dependence,\cite{Blazej2007} }
which is also shown by the negligible functional dependence of the Ti free energy \cite{Supplmental_Note:PRB}.
}
Because the LDA functional has been shown to be highly accurate for
describing the lattice dynamical and thermodynamic properties of both
pure Ti and TiO$_2$,\cite{Mei2009,*Hu2010,*Mei2011} we have selected to also use it here when including vibrational 
contributions to the free energy.
\textcolor{black}{
We also note that we only consider free energies at room temperature, where both the small thermal expansion 
of a solid may be safely omitted and the temperature-dependent free energies are well
described by harmonic phonons at fixed volumes (\emph{e.g.}, DFT
equilibrium volumes). The accuracy of the theoretical harmonic LDA free energies
are 
\textcolor{black}{compared to experiment for Ti and TiO$_2$ in Ref.~\onlinecite{Supplmental_Note:PRB}.}
}

\subsection{Thermodynamics}

The formation energy ($E_f$) for a Ti oxide (\emph{e.g.}, Ti$_m$O$_n$) is calculated
as
\begin{equation}{\label{Equ_formation_energy}}
E_f=E_{e}({\text{Ti}}_m{\text{O}}_n)-m{\cdot}E_{e}({\text{Ti}})-\frac{n}{2}E_e({\text{O}}_2)\,,
\end{equation}
where $E_e$ is the electronic total energy for the oxide,
pure Ti metal in the HCP structure, and an isolated O$_2$
molecule, respectively, obtained from DFT.
The calculated $E_f$'s for various Ti oxides
using different functionals are shown in \autoref{Table_Formation_Energies}.

\begin{figure*}
\centering
\scalebox{1.1}[1.1]{\includegraphics{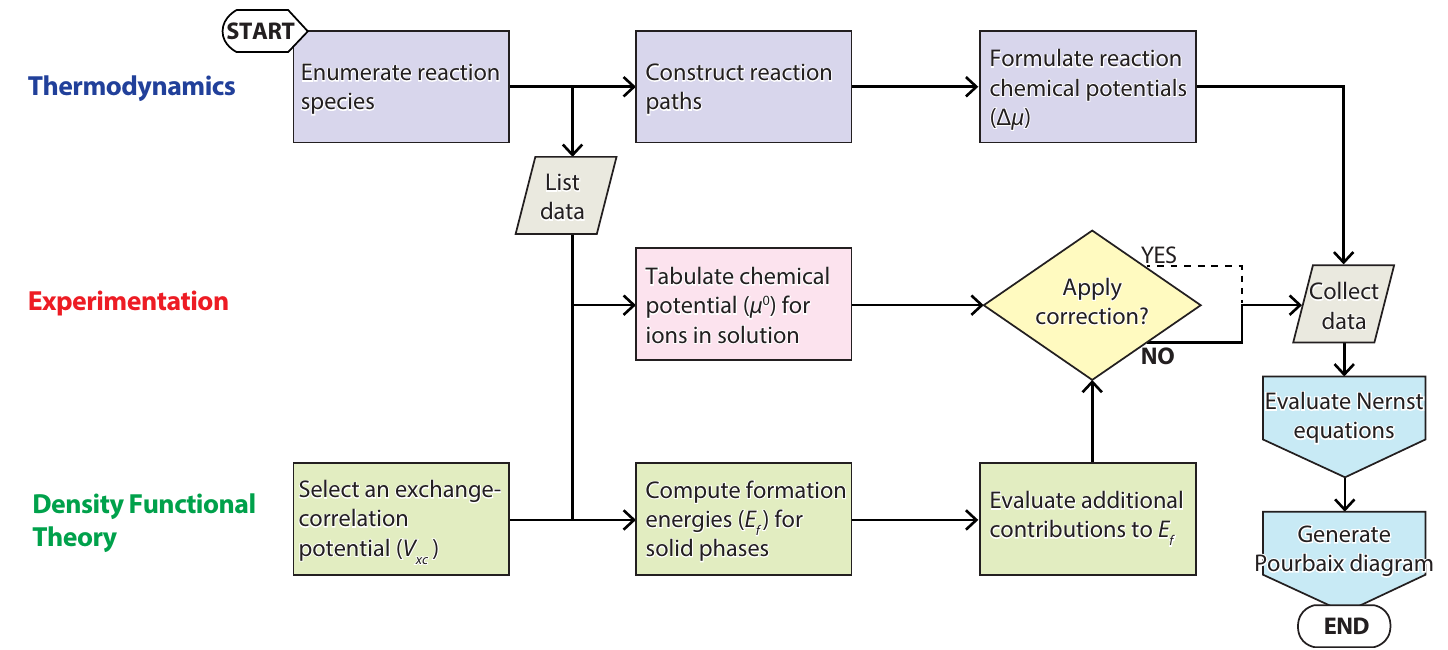}}\vspace{-0.5\baselineskip}
\caption{\label{Fig_work_flow} Workflow diagram for accurate
(correction free) simulation of electrochemical Pourbaix diagrams
from density functional theory. The process involves inputs from
thermodynamic formula, experimental data, and accurate density
functional theory. The path taken in this work follows the bolded
arrows and further details for what occurs at each step are given in
the main text. Of note, we point out that a variety of corrections
can be made to improve the agreement between density functional
theory values and experiment; however, we avoid these \emph{ad hoc} (and
often tedious) corrections by using more accurate functionals that
include additional forms of electronic exchange. In the outcome step
involving the Nernst equations, the chemical potential is computed
with collected theoretical and experimental data for each species
with respect to pH and electrode potential; the molar concentration
of the aqueous ions in solution is also specified. In the final
step, the lowest energy species per pH value and electrode potential
is obtained on a dense grid to generate the resulting Pourbaix
diagrams that appear throughout the text.}
\end{figure*}

When simulating the electrochemical
stability at room temperature (298.15 K), thermal contributions
should be included for the chemical potential of the oxides as
\begin{equation}{\label{Equ_oxide_chemical_potential}}
\mu({\text{Ti}}_m{\text{O}}_n)=E_f+{\Delta}F_{vib}-\frac{n}{2}{\Delta}F_T({\text{O}}_2)\,,
\end{equation}
where ${\Delta}F_{vib}$ is the contribution from the vibrational
free energies of  the Ti oxides (\autoref{Table_Formation_Energies}) and pure metal
[0.04 (0\,K) and -0.01 (298.15\,K) eV per Ti, respectively].
${\Delta}F_T({\text{O}}_2)$ is the free energy of
O$_2$ gas at the standard state (298.15 K and 1.0 bar),
which includes the contributions from the translation, rotation, and vibration of
the O$_2$ molecule.
${\Delta}F_T({\text{O}}_2)$ is 0.10 and -0.45 eV at 0 and 298.15 K,
respectively.
Often ${\Delta}F_{vib}$ is omitted in DFT simulations examining oxide stability
owing to its expensive computational costs, however, we explicitly include it
in this work.

Directly calculating the chemical
potential of ions in solution is still challenging for DFT, however,
the experimental standard chemical potentials ($\mu^0$; 298.15 K and
1.0 mol/L) of various ions are well established. Ti$^{++}$,
Ti$^{+++}$, (TiO)$^{++}$, and (HTiO$_3$)$^-$ are the most frequently
observed Ti ions in solution;  we have collected the chemical potentials for these ions
from experimental databases in \autoref{Tab_ion_energies}. Note that the
chemical potential of an aqueous ion (\emph{e.g.}, ion I) at a specified
concentration ([I]) can be expressed as
\begin{equation}{\label{Equ_ion_chemical_potential}}
\mu({\text{I}})=\mu^0+RT\log({\text{[I]}})\,,
\end{equation}
where $R$ is the gas constant [8.31446 J/(mol$\cdot$K)]. In
addition, the experimental $\mu^0$ for water is -2.458 eV.\cite{Chase1998}

\begin{table}
\caption{\label{Table_Formation_Energies} Calculated formation
energy ($E_f$) and vibrational free energy ($F_{vib}$) for Ti
oxides. $E_f$ is calculated using different density functionals,
while $F_{vib}$ is only calculated using LDA. $F_{vib}$ is absent
for TiO and Ti$_2$O$_7$, indicated by `---', owing to their dynamical instability at the LDA level.
\textcolor{black}{Note that there is a negligible functional dependence ($\lesssim$1.6 meV per Ti atom) 
up to room temperature on the vibrational energies \cite{Supplmental_Note:PRB}; an error of $<50$ meV per Ti atom is unlikely to have a substantial influence on the calculated Pourbaix diagrams.
}
}
\begin{ruledtabular}
\begin{tabular}{lccccccc}
            &\multicolumn{5}{c}{$E_f$ (eV/f.u.)}
            &\multicolumn{2}{c}{$F_{vib}$ (eV/f.u.)}\\
            \cline{2-6}\cline{7-8}\\[-0.5em]
%Species
        & LDA    & PBE    & PBEsol & MS2    & HSE06  & 0 K &
298.15 K \\\hline
Ti$_6$O     & -6.64 & -5.69 & -6.06 & -5.93 & -6.33 &0.34 &0.11 \\
Ti$_3$O     & -6.49 & -5.54 & -5.90 & -5.76 & -5.99 &0.23 &0.12 \\
Ti$_2$O     & -6.30 & -5.36 & -5.70 & -5.56 & -5.65 &0.18 &0.11 \\
Ti$_3$O$_2$ &-12.23 &-10.18 &-10.95 &-10.39 &-10.69 &0.34 &0.23 \\
TiO         & -5.51 & -4.52 & -4.87 & -4.50 & -4.46 & --- & --- \\
Ti$_2$O$_3$ &-16.46 &-14.28 &-14.90 &-14.62 &-14.99 &0.35 &0.26 \\
Ti$_3$O$_5$ &-26.68 &-23.35 &-24.25 &-24.01 &\textcolor{black}{-24.66} &0.58 &0.42 \\
Ti$_4$O$_7$ &-36.85 &-32.32 &-33.50 &-33.34 &\textcolor{black}{-34.03} &0.75 &0.51 \\
TiO$_2$     &-10.14 & -8.93 & -9.23 & -9.32 & -9.43 &0.21 &0.14 \\
Ti$_2$O$_7$ &-18.66 &-16.43 &-16.72 &-16.00 &-16.51 & --- &---  \\
\end{tabular}
\end{ruledtabular}
\end{table}

\begin{table}
\caption{\label{Tab_ion_energies} Standard chemical potential
($\mu^0$, in eV) of aqueous ions at 298.15 K obtained from four
experimental databases. The Pourbaix \& Bard's data (column 1) are used in
this work. Burgess's and Lide's data are calculated from the
redox potentials listed in the provided references.}
\begin{ruledtabular}
\begin{tabular}{lccc}
ion              &  Pourbaix \& Bard\cite{Pourbaix1966,Bard1985} &
Burgess\cite{Burgess1978} & Lide\cite{Lide2010}
\\\hline
Ti$^{++}$        &  -3.26    & -3.26     & -3.26 \\
Ti$^{+++}$       &  -3.63    & -3.63     & -4.16 \\
(TiO$_2$)$^{++}$ &  -4.84    & ---       &  ---  \\
(HTiO$_3$)$^-$   &  -9.91    & ---       &  ---  \\
(TiO)$^{++}$     &  -5.98    & ---       &  ---  \\
\end{tabular}
\end{ruledtabular}
\end{table}

\subsection{Simulation Method for Pourbaix Diagram}
The general workflow for our ab-initio simulation of the electrochemical Pourbaix diagrams
is shown in \autoref{Fig_work_flow}.
It involves the use of thermodynamic principles, data from experimentation, and
computed quantities from an electronic structure method.
%, which also can be readily transferred to calculate the Pourbaix diagram of other materials.
%
The involved steps are described in detail below using the Ti materials system, but
the approach is readily transferred to other material families:
\begin{enumerate}[(1)]
\item The workflow starts at enumerating involved reaction species, \emph{e.g.},
metal, oxides, water, and aqueous ions, which is indicated by
the first box in the `Thermodynamics' row of \autoref{Fig_work_flow}.

\item Knowing the involved species, we construct the reaction
paths that connect all of the species, % [box of "Construct reaction paths"],
and then derive the expressions of the reaction chemical
potentials for these paths [see the `Formulate reaction chemical
potentials ($\Delta\mu$)' box and \autoref{Append_Reactions}].

\item On the other hand, the required solution data, \emph{i.e.},
ion chemical potentials, are aggregated upon knowing the involved species (see the `List data' box).
Here the experimental standard chemical potentials  ($\mu^0$) of aqueous ions
in solution are used and tabulated in a subsequent step (\autoref{Tab_ion_energies}).

\item Next, density functional theory is used to compute the formation energies
of the Ti oxides [see `Compute $E_f$ for solid phases' boxes], before which one selects an exchange-correlation functional ($V_{xc}$) for use in the DFT calculation.
Importantly, the accuracy of the ab-initio Pourbaix diagrams is mainly determined by the performance of $V_{xc}$.

\item Having the calculated oxide formation energies, it may be necessary to
consider other additional contributions, \emph{e.g.}, vibrations, defects,
and impurities [`Evaluate additional contributions to
$E_f$' box]. Ti oxides can be readily contaminated by other elements
(\emph{e.g.}, Fe and C)
\cite{Mackey1994,Bartkowski1997,Baubande2002,*Hanaor2011}, and there
are always abundant spontaneous nonstoichiometric and stoichiometric
defects in Ti oxides, especially in TiO$_x$ with $x=0.5\sim2.0$
\cite{Grant1959,Murray1987,Leung1996,Bartkowski1997,Janotti2010,Watanabe1967,*Eriksson1993,*Waldner1999,*Cancarevic2007,*Jiang2013,*Jiang2013_2}.
The impurities and defects may considerably influence the structure
and stability \cite{Murray1987,Leung1996,Janotti2010,Jiang2013_2} of
Ti oxides. To avoid the complexity in doped and defective Ti oxides,
these two contributions are not considered here. In contrast, vibrational contributions to the formation energy are explicitly included.

\item After obtaining the chemical potentials for the
aqueous ions (Step 3) and oxide formation energies (Step 4 and 5),
the calculated formation energies (and/or sometimes the experimental $\mu^0$) 
may be corrected\cite{Persson2012} to obtain an accurate
dissolution energy for oxides (see the `Apply Correction'
decision box).
This kind of correction originates from the
inaccuracy of the DFT calculation owing to the selected
$V_{xc}$ in Step 4. Here, we avoid
using this artificial correction by using state-of-art density
functionals that include additional forms of electronic exchange (see bold path labeled
by `No').

\item %The experimental $\mu^0$'s of aqueous ions and calculated $E_f$'s of solid oxides are collected [box of "Collect data"],
All thermo-electrochemical data are then collected and the $\Delta\mu$
for each reaction path is obtained by evaluating the corresponding Nernst
expression at specified pH values and electrode potential $V$.
%[box of "Evaluate Nernst equations"].

\item To generate a Pourbaix diagram, a dense numerical grid
($4000\times4000$, here) spanning
the concerned two-dimensional pH--$V$ phase space (pH\,$\in[-2,16]$,
$V\,\in[-3,3]$ V here) is used, and the relative chemical potentials
of all the species with respect to a reference species (\emph{e.g.}, pure
Ti) are calculated at all grid points ($1.6\times10^7$ points,
here).
This procedure allows us to identify the equilibrium state on each grid
point and to accurately trace the evolution of the phase boundaries.
Finally, a Pourbaix diagram is generated after scanning the complete
numerical grid.

\end{enumerate}

\section{Results and Discussion}

\subsection{Functional Dependence of Titanate Formation Energies}
\begin{figure}
\centering
\includegraphics[width=\columnwidth]{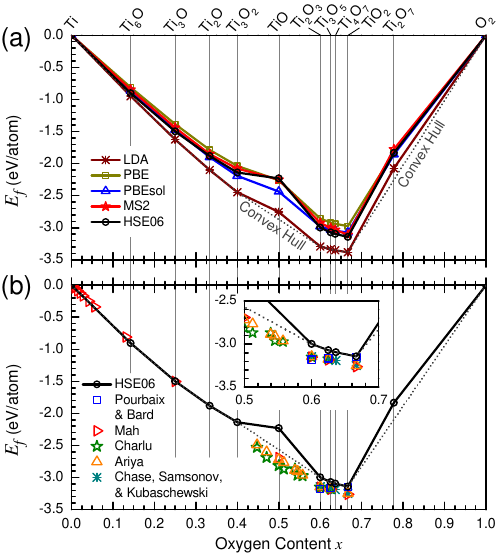}\vspace{-0.5\baselineskip}
\caption{\label{Fig_Formation_Energy} (a) Calculated formation
energies ($E_f$) of Ti oxides with respect to various
exchange-correlation potentials. (b) The comparison between HSE06
and experimental
results.\cite{,Pourbaix1966,Bard1985,Chase1998,*Humphrey1951,*Mah1955,*Ariya1957,*Charlu1974,*Samsonov1973,*Kubaschewski1993}
The inset in panel (b) shows in detail the range $x\in(0.5,0.7)$.}
\end{figure}

Owing to the importance of the exchange-correlation functional in accurately
reproducing experimental formation energies of solids, we first consider in detail the
electronic origin of the functionals' performance on Ti oxides.
The calculated formation energies ($E_f$) per formula unit for the Ti oxides
obtained with the LDA, PBE, PBEsol, MS2, and HSE06 functionals,
%as well as the vibrational free energies ($F_{vib}$) at 0 and 298.15 K from LDA functional,
are listed in \autoref{Table_Formation_Energies}.
To enable a direct comparison, these formation energies per atom are
plotted in \autoref{Fig_Formation_Energy}(a), and a comparison between the
values of $E_f$ obtained with the HSE06 functional and available
experimental values is given in panel (b).

First, we consider the general variation in formation energies obtained with local
exchange-correlation functionals.
The LDA functional \cite{Perdew1981} uses
the exchange-correlation potential of homogeneous electron gas, and
has the famous overbinding problem for formation/binding energies of solids
\cite{Perdew1996,Adllan2011,Jones1989}. Thus, the formation energies for all of the 
Ti oxides obtained from the LDA are the lowest among all the 
calculated results [\autoref{Table_Formation_Energies} and \autoref{Fig_Formation_Energy}(a)].
The PBE functional\cite{Perdew1992,Perdew1996,Perdew1997}, which includes additional
contributions of the electron density gradient, softens the bond strength
between ions with respect to that predicted by the LDA
\cite{Perdew1996}. Accordingly, the stability of the Ti oxides from PBE is 12-18\% lower than that from LDA.
The  PBEsol functional \cite{Perdew2008,*Perdew2009} enhances the stability of Ti oxides by 2-8\%
with respect to PBE (see \autoref{Table_Formation_Energies}), because the overbinding character of LDA is partially 
included in PBEsol.
Next, we consider the nonsemilocal functionals. 
%The metaGGA MS2 functional augments the exchange potential by including
%the electronic kinetic energy, which is also expressed
%as a density-gradient functional \cite{Sun2013,Sun2013_2}.
%%
%The hybrid HSE06 functional directly replaces 25\% of the semilocal
%exchange in PBE functional by the exact-nonlocal exchange with a screening length of within 10 \AA{}.
%\cite{Heyd2003,Heyd2004,Heyd2004_2,Heyd2006,Vydrov2006}
%
Although, PBEsol, MS2, and HSE06 functionals utilize different approaches
to improve the accuracy of electronic exchange, all approximations decrease the
formation energies for the Ti oxides with the exception that the values for TiO
are larger from MS2 and HSE06 (\autoref{Table_Formation_Energies}).
For example, the difference between the PBE
$E_f$(TiO) value and the corresponding convex hull is 0.2 eV/atom, which is
significantly increased to be $\gtrsim\!0.3$ eV/atom by the MS2 and HSE06
functionals, but decreased to be $\sim\!0.1$ eV by the PBEsol functional (LDA functional) [\autoref{Fig_Formation_Energy}(a)].
Since the electronic exchange in the PBEsol functional \cite{Perdew2008} is
designed to be close to that of the LDA, the relative stability between
different oxides obtained from the PBEsol functional is expected to differ
from that predicted with the MS2 and HSE06 functionals.
Now we examine the variation in the formation energy of select oxides.
The formation energies of Ti$_2$O$_7$ and TiO are clearly above the convex hull
(\autoref{Fig_Formation_Energy}),
indicating their thermodynamic instability.
In addition, they also
have imaginary phonon modes,\cite{Supplmental_Note:PRB} indicating
their dynamical instability.
\textcolor{black}{The instability of Ti$_2$O$_7$ ($x=0.78$) at the DFT level is understandable
owing to the highest experimentally observed O content phase
corresponding to $x=0.67$}.\cite{Murray1987}
On the other hand, TiO has been widely measured in high-temperature oxidation
experiments, and the experimental $E_f$(TiO) is quite close to the
convex hull [\autoref{Fig_Formation_Energy}(b)].
We note that in our calculations for TiO, we have used a pristine cubic rock salt structure; 
however, TiO is highly prone to
stoichiometric/nonstoichiometric defects (even up to 25 atomic percent
\cite{Bartkowski1997}), which transforms the ideal cubic TiO into
the experimentally measured monoclinic phase 
\cite{Gilles1969,Watanabe1967,Bartkowski1997,Leung1996}.
It is these defects that
stabilize the monoclinic TiO \cite{Leung1996,Jiang2013_2}, and
explain the existence of a `stable' TiO.
To avoid the complex defect structures in TiO, we treat only the pristine case,
and because it is not observed in electrochemical measurement at room temperature
\cite{Kelly1982}, where the stabilizing defects are difficult to form, we
omit it as a candidate phase in the electrochemical phase diagrams below.

\begin{figure}
\centering
%\scalebox{1.2}[1.2]{\includegraphics{LFHuang_fig4a.pdf}}\\
%\scalebox{1.2}[1.2]{\includegraphics{LFHuang_fig4b.pdf}}\\
%\scalebox{1.2}[1.2]{\includegraphics{LFHuang_fig4c.pdf}}\\
%\scalebox{1.2}[1.2]{\includegraphics{LFHuang_fig4d.pdf}}
{\includegraphics[width=0.9\columnwidth]{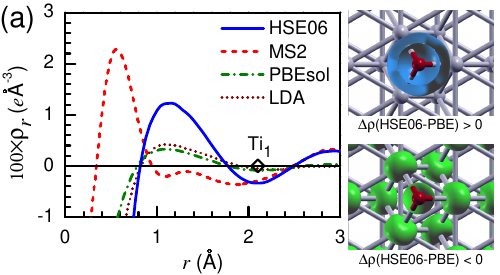}}\\
{\includegraphics[width=0.9\columnwidth]{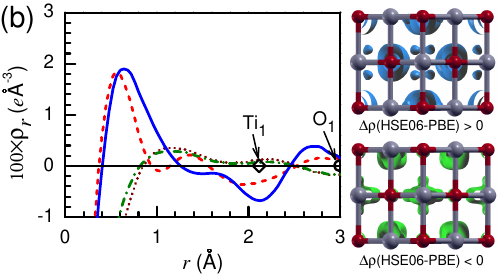}}\\
{\includegraphics[width=0.9\columnwidth]{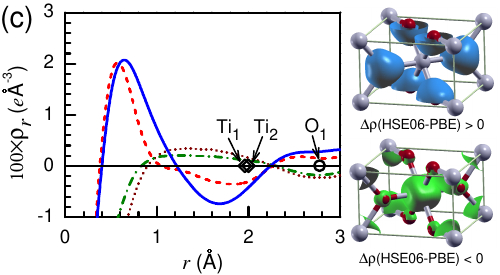}}\\
{\includegraphics[width=0.9\columnwidth]{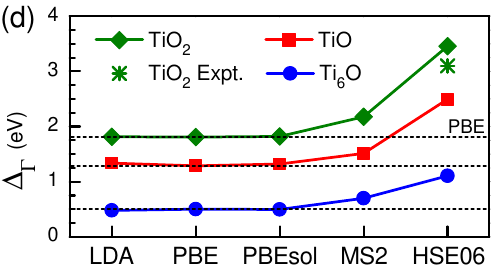}}
\caption{\label{Fig_diff_Q} %(a--c)
Functional-dependent radial
electron density ($\rho_r$) around an O atom (reference: PBE
$\rho_r$) and the electron density difference
$\Delta{\rho}$(HSE06-PBE) (isovalue: $\pm0.01$ $e$\AA{}$^{-3}$) in
(a) Ti$_6$O, (b) TiO, and (c) TiO$_2$, where the locations of the Ti
and O neighbors for an O atom located at the origin (0\,\AA) are indicated by hollow
diamonds and circles, respectively. (d) The functional-dependent
$\Gamma$-point band gap ($\Delta_\Gamma$) in Ti$_6$O, TiO, and
TiO$_2$, where the PBE gaps are indicated by dashed horizontal
lines, and the experimental $\Delta_\Gamma$ for rutile
TiO$_2$\cite{Grant1959,Samsonov1973,Amtout1995,Janotti2010} is
indicated by the star.}
\end{figure}

Defects are also widely
observed and have a sensitive temperature dependence in the Magn\'eli
phases (%Ti$_n$O$_{2n-1}$, e.g.,
Ti$_2$O$_3$, Ti$_3$O$_5$, and
Ti$_4$O$_7$) and TiO$_2$.
The former phases also can be viewed as Ti oxides with a different
density of extended defects \cite{Murray1987}, which
%Changeable defect concentration
results in
a continuous variation of $x$ from 0.40 to 0.67 as observed in
experimental samples.% of Magneli phases and TiO$_2$
\cite{Ariya1957,Merritt1973,Charlu1974,Murray1987,Bartkowski1997}
 Indeed,  the calculated HSE06 formation energies for
Ti$_2$O$_3$, Ti$_3$O$_5$, Ti$_4$O$_7$, and TiO$_2$ are higher than
the experimental ones by $\lesssim0.1$ eV per atom [\autoref{Fig_Formation_Energy}(b)]. 
This discrepancy is likely a result of the high temperatures
measurements. 
%, where the contribution from defects, as well as from their temperature dependence, to this theory-experiment discrepancy is still unknown.
%
%
%An additional investigation in the future is necessary to solve this
%issue.
Lastly, for the interstitial oxides with $x\leqslant0.25$, the
discrepancy between the convex hull and experimental formation energies
at the HSE 06 level
is only $0.00\sim0.02$ eV per atom, indicating the high accuracy of
the hybrid functional.

\begin{figure*}
\centering
\scalebox{1.1}[1.1]{\includegraphics{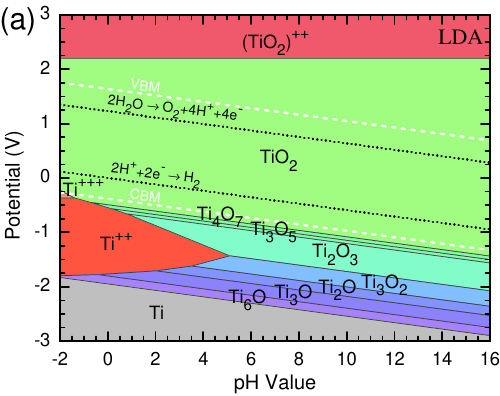}}
\scalebox{1.1}[1.1]{\includegraphics{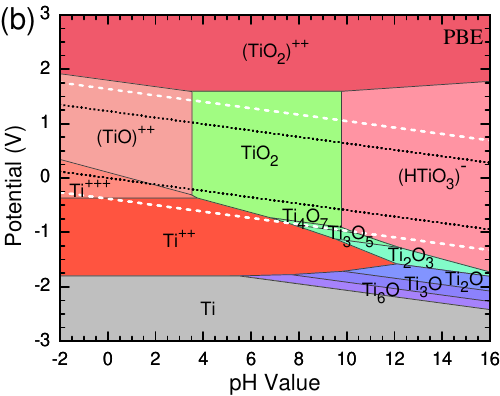}}
\scalebox{1.1}[1.1]{\includegraphics{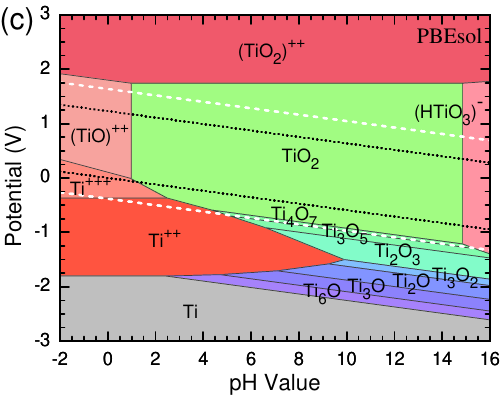}}\\
\scalebox{1.1}[1.1]{\includegraphics{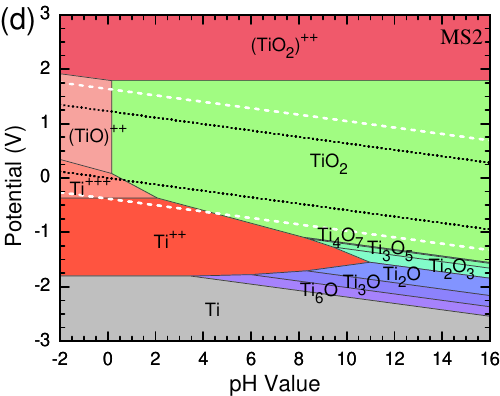}}
\scalebox{1.1}[1.1]{\includegraphics{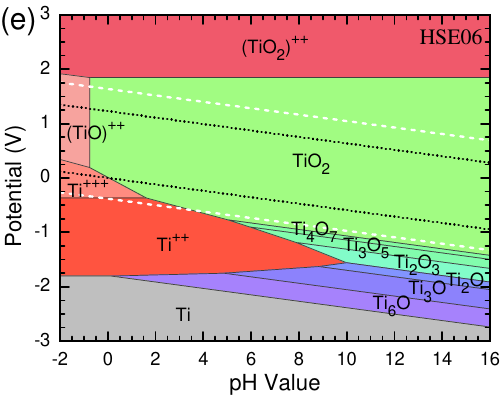}}
\scalebox{1.1}[1.1]{\includegraphics{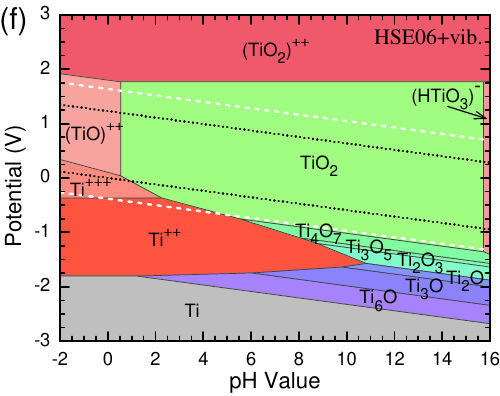}}\vspace{-0.5\baselineskip}
\caption{\label{Fig_Pourbaix_functionals} Pourbaix diagrams
calculated using (a) LDA, (b) PBE, (c) PBEsol, (d) MS2, and (e)
HSE06 functionals, as well as (f) HSE06 functional plus vibrational (vib) 
contributions. The CBM and VBM levels of rutile TiO$_2$, and the
electrode potentials for the oxidation of H$_2$O to O$_2$
(2H$_2$O${\rightarrow}$O$_2$+4H$^+$+4$e^-$) and for the reduction of
H$^+$ to H$_2$ (2H$^+$+2$e^-$${\rightarrow}$H$_2$) are obtained from
Ref.~\onlinecite{Gerischer1977}.}
\end{figure*}

%%%%
\subsection{Electronic Structure Dependencies}
To understand the electronic scale origin of the functional dependence in
more detail, we examine the radial electron density defined as
\begin{equation}{\label{Equ_Radius_rho}}
\rho_r=\frac{1}{4{\pi}r^2}\int_{|{\bf{r}}|=r}{\rho({\bf{r}})d{\bf{r}}}\,,
\end{equation}
where ${\bf{r}}$ is the position vector ($r$ is its length), and
$\rho({\bf{r}})$ is the electron density at the ${\bf{r}}$ point.
$\rho_r$ is the average electron density on a spherical surface with
radius of $r$. 
Throughout we take the position of an O atom as the origin.
According to the O concentration and lattice type, the interstital oxides (\emph{e.g.}, Ti$_6$O, Ti$_3$O, Ti$_2$O, and Ti$_3$O$_2$ here) 
and octahedral oxides (\emph{e.g.}, Ti$_2$O$_3$, Ti$_3$O$_5$, Ti$_4$O$_7$, TiO$_2$ here) 
are separated by rock salt TiO. 
To derive a generic electronic structure mechanism, Ti$_6$O, TiO, and TiO$_2$ are 
then selected to analyze the functional dependence of $\rho_r$, with each serving as 
a prototypical member of the interstitial, octahedral, and boundary oxides, respectively.
The calculated radial electron densities for Ti$_6$O, TiO, and TiO$_2$ are
shown in \autoref{Fig_diff_Q}(a--c), accompanied by the three-dimensional
electron density difference maps between PBE and HSE06, 
$\Delta\rho$(HSE06-PBE). We find from $\rho_r$ and $\Delta\rho$(HSE06-PBE) that,
compared with the PBE functional, the MS2 and HSE06 functionals
draw more electrons from both O and Ti atoms into an
interstitial shell region surrounding the O atom, which is expected to
increase the bond covalency and bond strength. For example in rutile TiO$_2$ [\autoref{Fig_diff_Q}(c)], 
the peak of $\rho_r$(HSE06-PBE) at 0.6 \AA{} and the corresponding positive shell in  
$\Delta\rho$(HSE06-PBE) both indicate the increase in electrons participating in the Ti-O bond by nonlocal 
electronic exchange. Interestingly, the PBEsol functional does not considerably increase 
the electron density along the bond (dash-dotted line);  its $\rho_r$ is nearly the same as that of the 
LDA functional (dotted line) as seen in \autoref{Fig_diff_Q}, because the LDA electronic exchange is partially used in the PBEsol functional.\cite{Perdew2008}
The increased bond covalency also can be reflected by the increase in band gap.
The band gap at the $\Gamma$ point [$\Delta_\Gamma$, \autoref{Fig_diff_Q}(d)] is
nearly the same from LDA, PBE, and PBEsol functional, due to the
similar bond covalency, however, it is observably higher at both the MS2
and HSE06 levels owing to the increased bond covalency. 
The phase stability always increases 
with bond covalency in covalent systems.
Therefore, we can conclude that the enhanced oxide stability obtained from 
the MS2 and HSE06 functional with respect to the PBE functional is mainly 
driven by the increased bond covalency, whereas the enhanced oxide stability 
from the PBEsol functional originates from the partial LDA over-binding character 
in PBEsol and not from a physically more precise electronic interaction.
\textcolor{black}{The former observation leads to a more accurate band gap of TiO$_2$ with the HSE06 functional 
compared to experiment [\autoref{Fig_diff_Q}(d)], 
and such functionals with exact 
exchange are likely to give more accurate physical properties.\cite{Franchini2005,Paier2006,Franchini2007,Hummer2007,Paier2008,Janotti2010} 
The latter then indicates} that the relative accuracy of PBEsol in the formation energies for Ti oxides are 
coincidental, and a systematic accuracy for PBEsol is not guaranteed. 
For example, PBEsol $E_b$(O$_2$) is inaccurate [\autoref{Fig_O2_Binding_Energy}(b)]. 
Thus, we conclude that the PBEsol functional should be used with care for transition-metal oxide
formation energies in the future.

\subsection{Electrochemical Phase Diagrams}
We now show how the changes in electronic structure alter the phase stability of
competing oxides phases, and in turn, modify the stability ranges in the
electrochemical diagrams.
The simulated Pourbaix diagrams using LDA, PBE, PBEsol, MS2,
and HSE06 functional, as well as HSE06 functional plus vibrational
contribution (HSE06$+$vib), are shown in \autoref{Fig_Pourbaix_functionals}.
Here, the ion concentration is set
to be a moderate value of $10^{-6}$ mol/L.
We also plot in each diagram the thermodynamic stability ranges for water, \emph{i.e.}, 
the phase space between its reduction (2H$^++2e^-\rightarrow$H$_2$) and 
oxidation (2H$_2$O$\rightarrow$O$_2$+4H$^+$+4$e^-$) boundaries. 
The valence band maximum (VBM) of rutile TiO$_2$ is also given. It 
is higher in potential than the water oxidation boundary, and its
conduction band minimum (CBM) is lower than the water reduction boundary. 
Thus, light-induced electron-hole pairs in TiO$_2$ can generate high enough voltages 
to decompose water (into H$_2$ and O$_2$ gases) or other organic 
molecules. This is the reason why TiO$_2$ is a promising photocatalyst for hydrogen production and 
pollutant elimination.\cite{Grant1959,Gerischer1977,Mo1995,Yigit2009,Hanaor2011,Persson2012}  
%%%

%%%
For all cases in \autoref{Fig_Pourbaix_functionals}, the phase space can be
divided into three kinds of domains \cite{Pourbaix1966}: ($i$)
\emph{immunity} (``I'') domain with pure Ti metal, ($ii$) \emph{passivation} (``P'')  domain with Ti
oxides, and ($iii$) \emph{corrosion} (``C'') domain with aqueous ions. The phase spaces for the considered species always 
vary with the type of used $V_{xc}$, although some general trends in these Pourbaix diagrams may be drawn: 
(1) As a base metal (not a noble metal), pure Ti is not stable at zero electrode potential, and the 
\emph{immunity} domain may only be accessed by a negative electrode potential (\emph{e.g.}, $<-1.84$ V); 
(2) Metal Ti will be directly corroded into aqueous ions in strong enough acid solutions, 
while in alkaline solutions, an oxide layer is expected to passivate the metal surface and 
protect the metal from corrosion; and (3) At high enough electrode potential (\emph{e.g.}, $>\!1.77$ V),
the aqueous ion (\emph{e.g.}, TiO$_2^{++}$) will be the preferred state. 
Note that the slopes of the phase boundaries are determined by the number of 
electrons and H$^+$ ions involved in the reaction (see \autoref{Append_Reactions} in the Appendix), 
and the boundary heights are determined by the 
formation energies.

Before discussing the results of our simulations, we first summarize relevant
electrochemical experiments as a guide to
which phases are stable under variable pH and potential; this understanding will then
be used to draw distinctions among the phase diagrams generated with
different exchange-correlation functionals.
TiO$_2$ is dissolved in acid solutions
with pH value of 0.3-0.8 \cite{Yin2001}, whereas it is stable in
solution with pH value of 2.0--11.0
\cite{Kroger2006,Hu2009,Yigit2009,Lee2014}.
Furthermore, Ti is spontaneously dissolved in solution when the
pH value is $<\!2.3-3.0$, at which an oxide
layer is assumed to exist and cover the Ti surface \cite{Kelly1982},
indicating that the passivating oxide layer starts to be dissolved
in strong acidic solutions and fails to protect the Ti metal.
Thus, the corrosion boundaries of TiO$_2$ should be at pH values of
$\sim$1.5 under acid conditions and $>\!$11.0 under alkaline condition,
respectively. Lastly, TiO$_2$ is stable in dilute acid
solutions; for example, an acidic solution with pH$\sim$4.5 is used in industry
to remove Fe compounds in rutile-structured minerals and synthesize
high-purity rutile TiO$_2$ products %with a high purity (e.g., $\gtrsim90$\%)
\cite{Mackey1994,Baubande2002}.

In the Ti Pourbaix diagram calculated at the LDA level
[\autoref{Fig_Pourbaix_functionals}(a)], the (TiO)$^{++}$ and
(HTiO$_3$)$^{-}$ aqueous ions are absent under acid and alkaline
conditions (pH$\in[-2,16]$), respectively, in constrast to experiment. 
The simulated TiO$_2$--Ti$^{+++}$ (at zero potnetial) and TiO$_2$--(TiO)$^{++}$ 
boundaries reside at pH values of less than -2.0, while the experimental dissolution boundary
of TiO$_2$ resides at a pH of $\sim$1.5.\cite{Kelly1982,Yin2001,Kroger2006,Yigit2009,Lee2014,Hu2009} 
This incorrect simulated phase stability is due to the overestimated stability of Ti
oxides by the LDA functional (\autoref{Fig_Formation_Energy}). 

In contrast, the PBE functional over-corrects that error, and
the corrosion boundaries for TiO$_2$ into the Ti$^{+++}$, (TiO)$^{++}$, and (HTiO$_3$)$^{-}$ aqueous ions are
at pH values of 2.1, 3.5, and 9.7, respectively, in the PBE Pourbaix diagram [\autoref{Fig_Pourbaix_functionals}(b)].
However, from the experimental observations,\cite{Kelly1982,Yin2001,Kroger2006,Yigit2009,Lee2014,Hu2009} 
TiO$_2$ should be stable within the pH range of $[2.0, 11.0]$. The phase spaces of other Ti oxides are also 
significantly reduced in the PBE Pourbaix diagram [\autoref{Fig_Pourbaix_functionals}(b)] 
compared with the LDA Pourbaix diagram [\autoref{Fig_Pourbaix_functionals}(a)].
The reduction in the phase stability of Ti oxides is driven by the
underestimated stability of Ti oxides by the PBE functional [\autoref{Fig_Formation_Energy}]. 
Owing to this oxide stability reduction by the PBE functional, the PBE \emph{immunity} and \emph{corrosion} domains are 
considerably larger than the LDA ones.
In an early simulated Pourbaix diagram of Ti that fully uses the experimental 
formation energies for oxides and aqueous ions,\cite{Pourbaix1966,Portero2011} 
the (TiO)$^{++}$ ion is absent in the Pourbaix diagram, and the TiO$_2$--Ti$^{+++}$ boundary at zero potential resides
at a pH values of approximately less than -1.0, which is also obviously incorrect---the 
experimental dissolution boundary $\sim1.5$\cite{Kelly1982,Yin2001,Kroger2006,Yigit2009,Lee2014,Hu2009}.
This incorrect prediction in electrochemical stability should be
ascribed to the direct usage of the oxide formation energies derived from
high-temperature combustion, 
\textcolor{black}{which include some contributions from abundant
defects.}

\begin{figure*}
\centering
\scalebox{1.05}[1.05]{\includegraphics{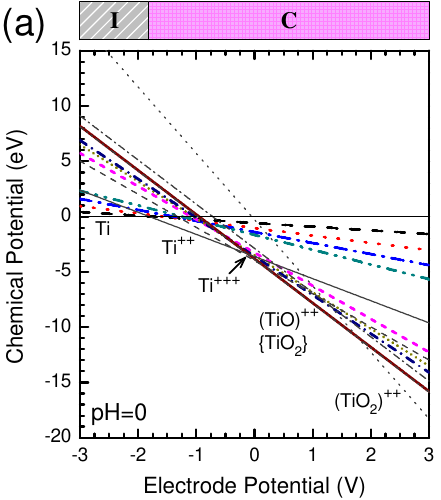}}
\scalebox{1.05}[1.05]{\includegraphics{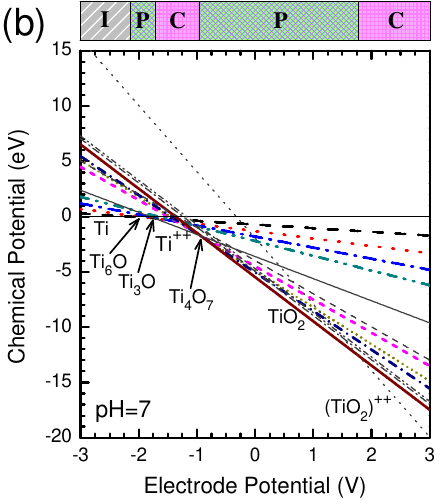}}
\scalebox{1.05}[1.05]{\includegraphics{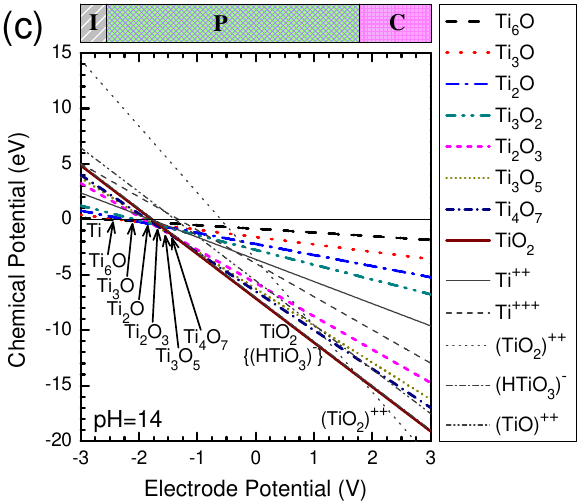}}\vspace{-0.5\baselineskip}
\caption{\label{Fig_Chemical_Potential} Variation of chemical potentials ($\mu$) with 
electrode potential at different pH values (0, 7, and 14) calculated from the 
HSE06$+$vib method, where $\mu(\text{Ti})$ is the reference. The species
with the lowest $\mu$ are labeled, and in panel (a) and (c), TiO$_2$
and (HTiO$_3$)$^-$ ion (in braces) are very close to the
lowest TiO$^{++}$ and (TiO$_2$)$^{++}$ ions, respectively. The
immunity (\textbf{I}), passivation (\textbf{P}), and corrosion (\textbf{C}) domains are
indicated on the panel top.}
\end{figure*}

%%%%
In the PBEsol, MS2, and HSE06 Pourbaix diagrams [\autoref{Fig_Pourbaix_functionals}(c)-(e)], the dissolution pH values for
TiO$_2$ into: 
($i$) Ti$^{+++}$ aqueous ion at zero electrode potential are 1.1, 0.5, and 0.0, respectively; 
($ii$) (TiO)$^{++}$ aqueous ions (at any electrode potential) are 1.0, 0.2, and -0.8, respectively; and 
($iii$) (HTiO$_3$)$^-$ aqueous ions (at any electrode potential) are $\gtrsim15.0$.
The electrochemical stability of TiO$_2$ in acidic solutions is well studied in experiment; thus, 
we focus on it here to evaluate the accuracy of these functionals.
From the acid dissolution boundary of TiO$_2$, these results are in better agreement with experiment and are an
improvement on the diagrams computed using the LDA and PBE
exchange-correlation functionals. 
Although, the (HTiO$_3$)$^-$ aqueous ion only appears in strong acidic 
solution for the PBEsol Pourbaix diagram, whereas it is absent in the MS2 and HSE06 Pourbaix diagrams, its observation
is supported by its precipitation mechanism (described below).

Comparing the HSE06 and HSE06$+$vib Pourbaix diagrams
[\autoref{Fig_Pourbaix_functionals}(e) and (f)], we find that including
the oxide vibrational free energy contributions shifts the
oxide corrosion boundaries [into Ti$^{++}$, Ti$^{+++}$, and (TiO)$^{++}$ aqueous ions] up to larger pH values by $\sim$1.0, 
and also makes (HTiO$_3$)$^-$ aqueous ion appear at strong alkaline condition in the Pourbaix diagram. 
The phase stability ranges of other Ti oxides are partially replaced by those of Ti$^{++}$ and pure Ti. 
The phonons have a negative contribution to the oxide stability, 
because the vibrational correction to the chemical potential 
($\Delta{F_{vib}}$ in \autoref{Equ_oxide_chemical_potential}) is positive (\autoref{Table_Formation_Energies}).
The vibrational correction shifts the pH values of the TiO$_2$--Ti$^{+++}$ and TiO$_2$--(TiO)$^{++}$ boundaries 
upwards by 0.7 and 1.3, respectively. 
Therefore, the vibration-corrected TiO$_2$--Ti$^{+++}$ boundary resides
at pH values of 1.8, 1.2, and 0.7 in the PBEsol, MS2, and HSE06 Pourbaix diagrams, respectively, and the 
vibration-corrected TiO$_2$--(TiO)$^{++}$ boundary resides
at pH values of 2.3, 1.5, and 0.5, respectively. 
All of these boundary pH values are very close to the
experimental dissolution boundary for TiO$_2$ mentioned above (pH$\sim$0.8--2.0).
Note that the accuracy achieved by the PBEsol functional in our Pourbaix diagram is 
accidental as described before. 
%due to the partially considered
%LDA over-binding character in PBEsol functional, while a systematic accuracy of PBEsol functional is not guaranteed, 
%\emph{e.g.}, the inaccuracy in $E_b$(O$_2$) as shown in \autoref{Fig_O2_Binding_Energy}. 
%
Because the 
\textcolor{black}{metaGGA} 
MS2 functional requires much less CPU time than a hybrid (HSE06) functional, 
but yields accurate $E_b$(O$_2$) (\autoref{Fig_O2_Binding_Energy}), oxide $E_f$'s (\autoref{Fig_Formation_Energy}), and therefore Pourbaix diagram (\autoref{Fig_Pourbaix_functionals}), we propose that it should be 
used for future calculations of non-magnetic materials when computational efficiency without \emph{ad hoc} corrections 
is required.

The variations of chemical
potentials ($\mu$) with electrode potential are shown in \autoref{Fig_Chemical_Potential},
where the pH values of 0, 7, and 14
are considered, respectively. These results are obtained from the HSE06$+$vib method, but note 
that similar conclusions for the phase stability trends 
are obtained from the other functionals. 
These variations of $\mu$ not only present the underlying energetic mechanism for the evolution 
in the phase stability described in the Pourbaix diagrams, but also
suggest the precipitation probability of some species. 
% The \emph{immunity} (``I''), \emph{corrosion} (``C''), and \emph{passivation} (``P'') domains are also indicated in the upper panels.
%
Upon decreasing the electrode potential, there is a well-defined stability order for the oxides at any pH value (from most to least stable):
\begin{eqnarray*}
\mathrm{TiO}_2\rightarrow \mathrm{Ti}_4\mathrm{O}_7&\rightarrow& \mathrm{Ti}_3\mathrm{O}_5\rightarrow \mathrm{Ti}_2\mathrm{O}_3\rightarrow\\\nonumber
&\rightarrow& \mathrm{Ti}_2\mathrm{O} (\gtrsim \mathrm{Ti}_3\mathrm{O}_2)\rightarrow \mathrm{Ti}_3\mathrm{O}\rightarrow \mathrm{Ti}_6\mathrm{O}\,,
\end{eqnarray*}
which is also observed in
the Pourbaix diagrams in \autoref{Fig_Pourbaix_functionals} by moving along the
axis of electrode potential.
This stability order for TiO$_2$ and the Magn\'eli phases is consistent
with electrochemical experiments, and among these phases, Ti$_2$O$_3$ also has been
measured to be the oxide stable at the lowest electrode potential 
\cite{Kelly1982}.

\begin{figure*}
\centering
\scalebox{1.1}[1.1]{\includegraphics{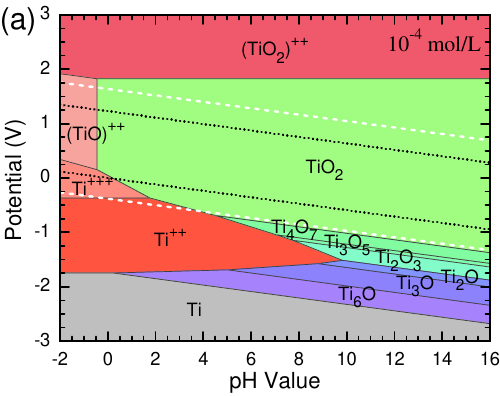}}
\scalebox{1.1}[1.1]{\includegraphics{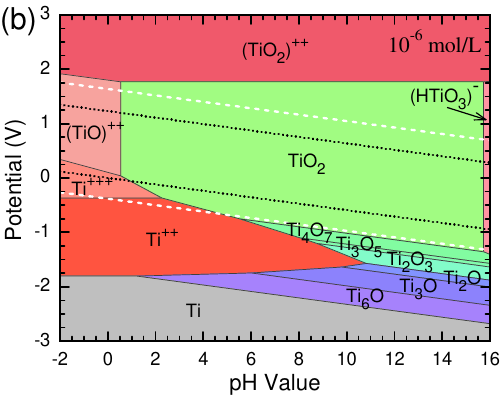}}
\scalebox{1.1}[1.1]{\includegraphics{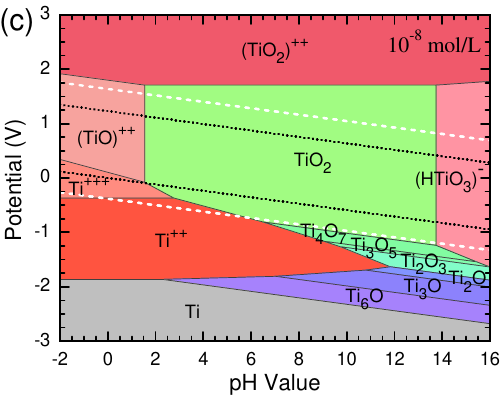}}\\
\scalebox{1.1}[1.1]{\includegraphics{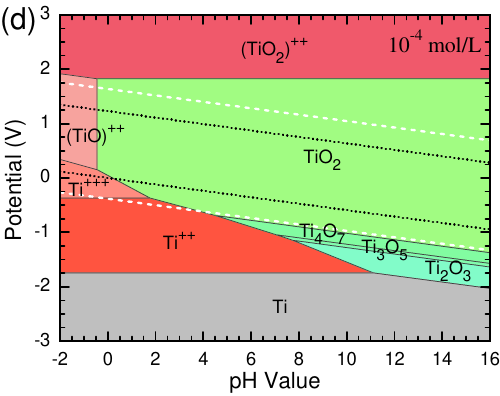}}
\scalebox{1.1}[1.1]{\includegraphics{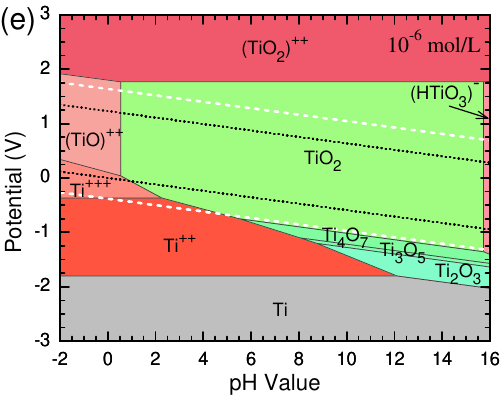}}
\scalebox{1.1}[1.1]{\includegraphics{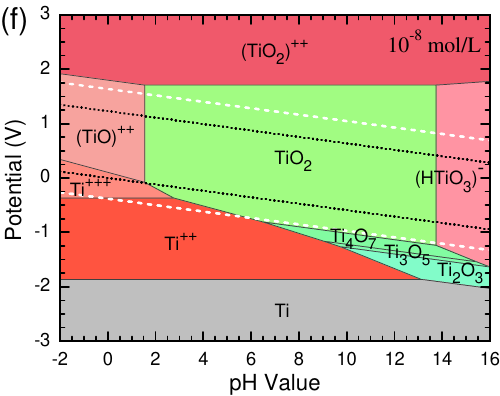}}\vspace{-0.5\baselineskip}
\caption{\label{Fig_Pourbaix_Ion_concentrations} Pourbaix diagrams
with aqueous ion concentrations of (a, d) $10^{-4}$, (b, e)
$10^{-6}$, and (c, f) $10^{-8}$ mol/L, obtained with the HSE06$+$vib method. The interstitial oxides are excluded in panels (d) through (e).}
\end{figure*}

Comparing the upper panels of \autoref{Fig_Chemical_Potential}, it can 
be explicitly seen that the \emph{passivation} domain expands with increasing pH value, because of
the downshift of oxide chemical potentials with respect to those of the aqueous ions and pure Ti. This 
oxide stabilization is due to the decreased number of H$^+$ ions available to react with Ti oxides (see Appendix).
Interestingly, the interstitial oxides have not been observed in
electrochemical experiments \cite{Kelly1982}.
This can be explained by recognizing that
the intercalation of O atoms into the HCP Ti lattice requires overcoming a
high diffusion barrier ($>$2.0\,eV) \cite{Henry2011}, and thus, there is likely insufficient
time for the ordered interstitial oxides to form on the surface of
metal Ti at room temperature.
(Pourbaix diagrams without the interstitial oxide phases are given in \autoref{Fig_Pourbaix_Ion_concentrations}.)
TiO also has not been observed in electrochemical experiments in
solutions at 298.15 K \cite{Kelly1982}.
As discussed above, TiO has
been widely observed in high-temperature phase diagrams (oxygen
content: $0.40\sim0.55$)
\cite{Gilles1969,Watanabe1967,Bartkowski1997,Leung1996}, owing to the
stabilization effect from the abundant defects formed at high
temperatures. At room temperature, there is no such defect formation
mechanism, which explains the absence of TiO in electrochemical
experiments.

In acidic solution with $\mathrm{pH}=0$ [\autoref{Fig_Chemical_Potential}(a)], Ti$^{++}$, Ti$^{+++}$,
(TiO)$^{++}$, and TiO$_2$ have very similar chemical potential values at the zero
electrode potential.
The latter two are also close for all electrode potential values.
These energetics can be used to justify why
amorphous TiO$_2$ controllably precipitates from a suspension
solution of TiCl$_4$ and Ti(OC$_2$H$_5$) at pH values
around $\sim$0.3 \cite{Yin2001}. Under acidic conditions,
TiO$_2$ is a little more stable than the Magn\'eli phases (Ti$_2$O$_3$, Ti$_3$O$_5$, and Ti$_4$O$_7$) at zero electrode potential,
which may be the reason why TiO$_2$ is
observed as an intermediate phase during the dissolution of
Ti$_3$O$_5$ in strong disulphuric solution \cite{Grey1988}.

With increasing pH value from 0 [\autoref{Fig_Chemical_Potential}(a)] to 7 [\autoref{Fig_Chemical_Potential}(b)],
metal Ti is more protected by the passivating
oxide, while only at a high electrode potential ($>1.77$ V), metal Ti or
Ti oxides will be corroded into (TiO$_2$)$^{++}$ aqueous ions.
This can be seen from the evolution in
the chemical potentials in \autoref{Fig_Chemical_Potential}(b), and in
the simplified stability regimes depicted in the upper panel.
In the alkaline solution with pH$=14$ [\autoref{Fig_Chemical_Potential}(c)],
metal Ti is highly passivated upon increasing the electrode potential
before being fully corroded into (TiO$_2$)$^{++}$ ions.

In the HSE06$+$vib Pourbaix diagram [\autoref{Fig_Pourbaix_functionals}(f)],
although the (HTiO$_3$)$^{-}$ aqueous ion only becomes the lowest
chemical potential specie at very strong alkaline conditions 
($\mathrm{pH}\gtrsim15.5$), it nonetheless can easily precipitate at a
relatively weaker alkaline condition (\emph{e.g.}, precipitation probability $\gtrsim10^{-3}$ at $\mathrm{pH}\gtrsim12.7$), 
because of its comparable chemical potential with respect to TiO$_2$ in 
alkaline solution [\autoref{Fig_Chemical_Potential}(c)]. This precipitation
mechanism should be responsible for the observation of aqueous 
(HTiO$_3$)$^{-}$  ions in experiment. 
Note that TiO$_2$ is found to be quite stable 
in alkaline solutions for pH values up to 11.0,\cite{Kelly1982,Yin2001,Kroger2006,Yigit2009,Lee2014,Hu2009}
because those solutions are still not alkaline enough for an obvious precipitation of aqueous ions.

%Having achieved highly accurate theoretical electrochemical
%stability of Ti and its oxides, 

We now examine the variation of the
Pourbaix diagrams on the aqueous-ion concentration [from $10^{-4}$
to $10^{-8}$ mol/L, \autoref{Fig_Pourbaix_Ion_concentrations}(a--c)].
A $10^{-4}$ mol/L solution has a relatively large aqueous ion concentration whereas 
%$10^{-6}$ mol/L is a moderate one; 
$10^{-8}$ mol/L corresponds to very high purity water, \emph{e.g.}, nuclear reactor water.\cite{Beverskog1997}
We find that upon decreasing the ion concentration by a factor of
$10^2$, the pH values for the TiO$_2$--Ti$^{++}$,
TiO$_2$--Ti$^{+++}$, TiO$_2$--(TiO)$^{++}$, and
TiO$_2$--(HTiO$_3$)$^{-}$ boundaries change by about 0.5, 0.5, 1.0,
and -2.0, respectively, because of the lower chemical potentials of
these species at lower concentration (see, \autoref{Equ_ion_chemical_potential}). However,
the electrode potentials for the dissolution (and passivation) boundaries
of Ti oxides (metal Ti) have no obvious variation ($<0.1$ V from
$10^{-4}$ to $10^{-8}$ mol/L).

Pourbaix diagrams omitting the interstitial oxides not observed
experimentally are shown in \autoref{Fig_Pourbaix_Ion_concentrations}(d--f).
The excluded passivation domains with interstitial oxides are mainly replaced
by the \emph{immunity} domain with metal Ti, and partially replaced by the
\emph{corrosion} domain with Ti$^{++}$ aqueous ion. 
The boundary pH values have similar variations with ion concentration as those 
described above when the interstitial oxides are explicitly included.
The Pourbaix diagrams with interstitial oxides [\autoref{Fig_Pourbaix_Ion_concentrations}(a--c)]
are useful in the experimental/theoretical studies when there is any factor (\emph{e.g.}, catalyst) that facilitates 
the oxygen diffusion in pure Ti and then promotes the formation of interstitial oxides on Ti surface;
while, the Pourbaix diagrams without interstitial oxides [\autoref{Fig_Pourbaix_Ion_concentrations}(d--f)] 
are ideally suited for conventional electrochemical studies.

\section{Conclusion}
We formulated an ab-initio workflow for accurate simulation of the
electrochemical phase equilibria under variable pH and potential, where 
vibrational effects are also taken into account. 
%When the simulation method
%was not accurate, some \emph{ad hoc} corrections were usually used 
%in the calculation of formation energies. However, stat-of-art DFT methods were 
%used to avoid such artificial corrections in a physical way, and the simulation 
%accuracy here was also well understood.
%
The protocol was applied to the Ti--O system, whereby the formation
energies of Ti oxides and Ti Pourbaix diagrams were
systematically calculated using semilocal (LDA, PBE, and PBEsol),
\textcolor{black}{metaGGA (MS2),} 
and hybrid (HSE06) density functionals. This comparative simulation approach uncovered the
correlation between the simulated electrochemical
stability accuracy (determined by consistency with experimental observations)
and the electronic-exchange precision in the exchange-correlation functional
used within the DFT framework.
The stability variations were then explained by the density-functional
dependent electronic structure in the oxides.

Various experimental phenomena that have been inconsistent with previous
calculated Ti Pourbaix diagrams were also explained by and become consistent
with our most accurate Pourbaix diagram obtained using the
HSE06 functional plus vibrational contributions.
Furthermore, we found that when the experimental
formation energies of Ti oxides estimated from combustion heats,
which include contributions from defects that inevitably form
at high temperatures, are directly used %using these archived data
to predict low-temperature Pourbaix diagrams, then significant inaccuracies are likely to result. 
Alternatively, the more accurate MS2 and HSE06 formation
energies can be used as the `true' formation energies for pristine Ti
oxides in the future.
These accurate ab-initio predicted Pourbaix diagrams obtained using
advanced exchange-correlation functionals without any \emph{ad hoc} corrections
are expected to be useful for both scientific and
industrial exploitation for the design of high-performing corossion resistant alloys, 
energy storage materials, and biocompatible implants.

%\section*{Acknowledgments}
\begin{acknowledgments}
L.-F.H.\ and J.M.R.\ wish to thank Prof.\ J.R.\ Scully (University of Virginia),  
Dr.\ Qimin Yan (Lawrence Berkeley National Laboratory), and 
Prof.\ L.D.\ Marks (Northwestern University) for insightful discussions. 
L.-F.H.\ and J.M.R.\ were
supported by the ONR MURI ``Understanding Atomic Scale Structure in
Four Dimensions to Design and Control Corrosion Resistant Alloys''
under Grant No.\ N00014-14-1-0675. Calculations were performed using
the QUEST HPC Facility at Northwestern University and the HPCMP
facilities at the Navy DoD Supercomputing Resource Center.

%which is jointly supported by the Office of the Provost, the Office
%for Research, and Northwestern University Information Technology.
%
%\jmr{Did you use any of the HPCMP facilities or STAMPEDE?}
\end{acknowledgments}

\bibliography{Reference_list}

%%%%%%%%%%%%%%%%%%%%%%%%%%%%%%%%%%%%%
%%%%%%%%%       Table       %%%%%%%%%
%%%%%%%%%%%%%%%%%%%%%%%%%%%%%%%%%%%%%

%%%%%%%%%%%%%%%%%%%%%%%%%%%%%%%%%%%%%
%%%%%%%%%       Figures     %%%%%%%%%
%%%%%%%%%%%%%%%%%%%%%%%%%%%%%%%%%%%%%

%\begin{equation}{\label{Equ_a_T}}
%\end{equation}

%\begin{table}[p]
%\caption{\label{}  }
%\begin{ruledtabular}
%\begin{tabular}{cccccc}
%
%\end{tabular}
%\end{ruledtabular}
%\end{table}

%\begin{figure}[p]
%\scalebox{1}[1]{\includegraphics{lfhuang_fig1.eps}}
%\caption{\label{}  }
%\end{figure}

%%%%%%%%%%%%%%%%%%%%%%%%%%%%%%%%%%%%%
%%%%%%%%%      APENDIX      %%%%%%%%%
%%%%%%%%%%%%%%%%%%%%%%%%%%%%%%%%%%%%%
\widetext
\appendix*

\section{Reaction paths and reaction energies}

In aqueous environments, the relative electrochemical stability between various species
(\emph{e.g.}, metal, oxides, and aqueous ions) are calculated from the
reaction chemical potentials ($\Delta\mu$) for the reactions paths
that connect all of them. The considered reaction paths used for the Ti--O system
and the associated reaction $\Delta\mu$'s are listed in \autoref{Append_Reactions}.

\begingroup
\squeezetable
\begin{table*}[h]
\caption{\label{Append_Reactions} Reaction paths and the
corresponding reaction energies ($\Delta\mu$, in kJ/mol), where
$\mu(\rm Ti)$ and $\mu(\rm H^+)$ are the references (\emph{i.e.}, zero) for the
chemical potentials at standard condition, and the standard hydrogen
potential is the reference for the electrode potential $U_p$ (in V).
$F$ is the Faraday constant
($=eN_A=9.65\times10^4$C$\cdot$mol$^{-1}$), and 1.0 kJ/mol$=F$ eV.}
\begin{ruledtabular}
\begin{tabular}{ll}
Reaction Path                 & $\Delta\mu$ \\\hline\\[-0.5em]
$\mathrm{Ti} \longrightarrow \mathrm{Ti}^{++}+2e^-$                      & $\Delta\mu(\mathrm{Ti}-\mathrm{Ti}^{++})=\mu(\mathrm{Ti})-\mu(\mathrm{Ti}^{++})+2FU=-\mu(\mathrm{Ti}^{++})+2FU_p$                          \\[0.5em]%
$\mathrm{Ti}^{+++}+e^- \longrightarrow \mathrm{Ti}^{++}$                 & $\Delta\mu(\mathrm{Ti}^{+++}-\mathrm{Ti}^{++})=\mu(\mathrm{Ti}^{+++})-\mu(\mathrm{Ti}^{++})-FU_p$                                 \\[0.5em] %
$(\mathrm{HTiO_3})^-+5\mathrm{H}^++2e^- \longrightarrow \mathrm{Ti}^{++}+3\mathrm{H_2O}$   & $\Delta\mu(\mathrm{HTiO_3}^--\mathrm{Ti}^{++})=\mu(\mathrm{HTiO_3}^-)-\mu(\mathrm{Ti}^{++})-3\mu(\mathrm{H_2O})-5RT\ln(10)\cdot{pH}-2FU_p$ \\[0.5em] %
$(\mathrm{TiO_2})^{++}+4\mathrm{H}^++4e^- \longrightarrow \mathrm{Ti}^{++}+2\mathrm{H_2O}$ & $\Delta\mu(\mathrm{TiO_2}^{++}-\mathrm{Ti}^{++})=\mu(\mathrm{TiO_2}^{++})-\mu(\mathrm{Ti}^{++})-4RT\ln(10)\cdot{pH}-4FU_p$        \\[0.5em] %
$(\mathrm{TiO})^{++}+2\mathrm{H}^++2e^- \longrightarrow \mathrm{Ti}^{++}+\mathrm{H_2O}$    & $\Delta\mu(\mathrm{TiO}^{++}-\mathrm{Ti}^{++})=\mu(\mathrm{TiO}^{++})-\mu(\mathrm{Ti}^{++})-\mu(\mathrm{H_2O})-2RT\ln(10)\cdot{pH}-2FU_p$  \\[0.5em] %
$\frac{1}{6}\mathrm{Ti_6O}+\frac{1}{3}\mathrm{H}^+ \longrightarrow \mathrm{Ti}^{++}+\frac{1}{6}\mathrm{H_2O}+\frac{5}{3}e^-$    & $\Delta\mu(\mathrm{Ti_6O}-\mathrm{Ti}^{++})=\frac{1}{6}\mu(\mathrm{Ti_6O})-\mu(\mathrm{Ti}^{++})-\frac{1}{6}\mu(\mathrm{H_2O})-\frac{1}{3}RT\ln(10)\cdot{pH}+\frac{5}{3}FU_p$      \\[0.5em] %
$\frac{1}{3}\mathrm{Ti_3O}+\frac{2}{3}\mathrm{H}^+ \longrightarrow \mathrm{Ti}^{++}+\frac{1}{3}\mathrm{H_2O}+\frac{4}{3}e^-$    & $\Delta\mu(\mathrm{Ti_3O}-\mathrm{Ti}^{++})=\frac{1}{3}\mu(\mathrm{Ti_3O})-\mu(\mathrm{Ti}^{++})-\frac{1}{3}\mu(\mathrm{H_2O})-\frac{2}{3}RT\ln(10)\cdot{pH}+\frac{4}{3}FU_p$      \\[0.5em] %
$\frac{1}{2}\mathrm{Ti_2O}+\mathrm{H}^+ \longrightarrow \mathrm{Ti}^{++}+\frac{1}{2}\mathrm{H_2O}+e^-$                          & $\Delta\mu(\mathrm{Ti_2O}-\mathrm{Ti}^{++})=\frac{1}{2}\mu(\mathrm{Ti_2O})-\mu(\mathrm{Ti}^{++})-\frac{1}{2}\mu(\mathrm{H_2O})-RT\ln(10)\cdot{pH}+FU_p$                            \\[0.5em] %
$\frac{1}{3}\mathrm{Ti_3O_2}+\frac{4}{3}\mathrm{H}^+ \longrightarrow \mathrm{Ti}^{++}+\frac{2}{3}\mathrm{H_2O}+\frac{2}{3}e^-$  & $\Delta\mu(\mathrm{Ti_3O_2}-\mathrm{Ti}^{++})=\frac{1}{3}\mu(\mathrm{Ti_3O_2})-\mu(\mathrm{Ti}^{++})-\frac{2}{3}\mu(\mathrm{H_2O})-\frac{4}{3}RT\ln(10)\cdot{pH}+\frac{2}{3}FU_p$  \\[0.5em] %
$\frac{1}{2}\mathrm{Ti_2O_3}+3\mathrm{H}^++e^- \longrightarrow \mathrm{Ti}^{++}+\frac{3}{2}\mathrm{H_2O}$                       & $\Delta\mu(\mathrm{Ti_2O_3}-\mathrm{Ti}^{++})=\frac{1}{2}\mu(\mathrm{Ti_2O_3})-\mu(\mathrm{Ti}^{++})-\frac{3}{2}\mu(\mathrm{H_2O})-3RT\ln(10)\cdot{pH}-FU_p$                       \\[0.5em] %
$\frac{1}{3}\mathrm{Ti_3O_5}+\frac{10}{3}\mathrm{H}^++\frac{4}{3}e^- \longrightarrow \mathrm{Ti}^{++}+\frac{5}{3}\mathrm{H_2O}$ & $\Delta\mu(\mathrm{Ti_3O_5}-\mathrm{Ti}^{++})=\frac{1}{3}\mu(\mathrm{Ti_3O_5})-\mu(\mathrm{Ti}^{++})-\frac{5}{3}\mu(\mathrm{H_2O})-\frac{10}{3}RT\ln(10)\cdot{pH}-\frac{4}{3}FU_p$ \\[0.5em] %
$\frac{1}{4}\mathrm{Ti}_4O_7+\frac{7}{2}\mathrm{H}^++\frac{3}{2}e^- \longrightarrow \mathrm{Ti}^{++}+\frac{7}{4}\mathrm{H_2O}$  & $\Delta\mu(\mathrm{Ti_4O_7}-\mathrm{Ti}^{++})=\frac{1}{4}\mu(\mathrm{Ti_4O_7})-\mu(\mathrm{Ti}^{++})-\frac{7}{4}\mu(\mathrm{H_2O})-\frac{7}{2}RT\ln(10)\cdot{pH}-\frac{3}{2}FU_p$  \\[0.5em] %
$\mathrm{TiO_2}+4\mathrm{H}^++2e^- \longrightarrow \mathrm{Ti}^{++}+2\mathrm{H_2O}$ & $\Delta\mu(\mathrm{TiO_2}-\mathrm{Ti}^{++})=\mu(\mathrm{TiO_2})-\mu(\mathrm{Ti}^{++})-2\mu(\mathrm{H_2O})-4RT\ln(10)\cdot{pH}-2FU_p$ \\ %
\end{tabular}
\end{ruledtabular}
\end{table*}
\endgroup

\end{document}